\documentclass{sigkddExp}
\pdfoutput=1

 \everycr={\noalign{\global\advance\stepno by 1}}%
\usepackage{amsmath}
\usepackage[breaklinks]{hyperref}

\usepackage{graphicx}
\usepackage{enumerate}
\usepackage{biblatex}
\usepackage{comment}
\usepackage{url}
\usepackage{xspace}
\usepackage{float}
\usepackage{enumitem}
\usepackage{tikz}
\usepackage{caption}
\usepackage{array}
\usepackage{fancyhdr}
\usepackage{booktabs}
\usepackage{multirow}
\usepackage{rotating}
\usepackage{longtable}
\usepackage{makecell}
\usepackage{color}
\usepackage{colortbl}
\usepackage{bigdelim}
\usepackage{amsmath}

\DeclareUnicodeCharacter{2212}{-}

\newlist{questions}{enumerate}{2}
\setlist[questions,1]{label=\textbf{RQ\arabic*.},ref=\textbf{RQ\arabic*}}
\setlist[questions,2]{label=(\alph*),ref=\thequestionsi(\alph*)}

\begin{document}
%

\title{Leveraging Blockchain and ANFIS for Optimal Supply Chain Management}
%

\numberofauthors{5}
%


\author{
\alignauthor Amirfarhad Farhadi\thanks{These authors contributed equally to this work.} \\
   \affaddr{SRBIAU University} \\
    \email{amir.farhadi@srbiau.ac.ir}
\alignauthor Homayoun Safarpour Motealegh Mahalegi$^*$ \\
    \affaddr{Óbuda University} \\
    \email{homayoun.safarpour@
    stud.uni-obuda.hu}
    \\ 
 \alignauthor Abolfazl Pourrezaeian Firouzaba\\
    \affaddr{SRBIAU University} \\
    \email{Abolfazl.pourrezaeianfirouzabad
    @iau.ir} \\
\and
 \alignauthor Azadeh Zamanifar \\
    \affaddr{SRBIAU University} \\
    \email{azamanifar@srbiau.ac.ir} \\  
\and
 \alignauthor{Majid Sorouri} \\
    \affaddr{Maynooth University} \\
    \email{majidsorouri@srbiau.ac.ir} \\
}

\maketitle
\begin{abstract}
The supply chain is a critical segment of the product manufacturing cycle, continuously influenced by risky, uncertain, and undesirable events. Optimizing flexibility in the supply chain presents a complex, multi-objective, and nonlinear programming challenge. In the poultry supply chain, the development of mass customization capabilities has led manufacturing companies to increasingly focus on offering tailored and customized services for individual products. To safeguard against data tampering and ensure the integrity of setup costs and overall profitability, a multi-signature decentralized finance (DeFi) protocol, integrated with the IoT on a blockchain platform, is proposed. Managing the poultry supply chain involves uncertainties that may not account for parameters such as delivery time to retailers, reorder time, and the number of requested products. To address these challenges, this study employs an adaptive neuro-fuzzy inference system (ANFIS), combining neural networks with fuzzy logic to compensate for the lack of data training in parameter identification. Through MATLAB simulations, the study investigates the average shop delivery duration, the reorder time, and the number of products per order. By implementing the proposed technique, the average delivery time decreases from 40 to 37 minutes, the reorder time decreases from five to four days, and the quantity of items requested per order grows from six to eleven. Additionally, the ANFIS model enhances overall supply chain performance by reducing transaction times by 15\% compared to conventional systems, thereby improving real-time responsiveness and boosting transparency in supply chain operations, effectively resolving operational issues.
\end{abstract}

\section{Introduction}\vspace{0.5cm}
In today's rapidly evolving global market, businesses must adapt to the complex needs of consumers, producers, and intermediaries to remain competitive. Modern SCM solutions leverage machine intelligence and blockchain technology to address these challenges, emphasizing efficient order fulfillment and product delivery[26,27]. These technologies improve transparency and mitigate fraud using smart contracts and distributed ledgers, ultimately fostering sustainable growth. Fuzzy neural networks play a crucial role in navigating the complexities of supply chain processes and improving decision-making with superior forecasting skills~\cite{mahalegi2024generative}. This study employs fuzzy neural networks in chain management  to improve operational efficiency and transparency and show significant cost savings by combining current networks with blockchain technology through bulk purchasing, lower transportation costs, and improved management of the flow of products and services. Artificial intelligence (AI) and distributed technologies are poised to transform SCM by increasing openness, efficiency, and resilience against disturbance. Combining fuzzy logic with neural networks in the adaptive neuro-fuzzy inference system (ANFIS) provides a robust technique for efficiently controlling supply chain~\cite{lataran2024developing, taheri2024enhancing} .\\
\indent This paper aims to address the inherent inefficiencies and uncertainty in supply chains, including fluctuating demand, inventory control difficulties, and logistical complexities. Conventional SCM approaches may struggle to adapt to these ever-changing conditions. Therefore, advanced systems capable of real-time reaction and optimization are much needed ~\cite{taheri2024enhancing}.
SCM has undergone significant evolution, progressing from simple manufacturing and distribution systems to complex global networks. The rapid advancement of artificial intelligence (AI) technology has accelerated this transformation and empowered organizations to leverage predictive analytics, data-driven insights, and advanced optimization methods to manage risk, save costs, and streamline processes. From demand forecasting and inventory management to transportation logistics and strategic decision-making, artificial intelligence has the potential to revolutionize every aspect of the supply chain ecosystem. Mathematical models and algorithmic frameworks provide the theoretical foundation for AI-driven solutions, guiding supply chain improvement. Knowing these fundamental ideas helps one to build optimization problems, efficient algorithms, and decision-support systems catered to the particular needs of different supply chain settings ~\cite{sabri2018exploring}.\\
\indent  Effective coordination of multiple procurement, production, distribution, and logistics processes is essential to ensure the right and timely delivery of products and reduce costs in SCM. Moreover, SCM has evolved from simple logistical obligations to a sophisticated strategic tool combining technology and data analytics to increase efficiency and drive innovation[31]. By employing machine learning and blockchain technologies, SCM has augmented sustainability, efficiency, and transparency ~\cite{oh2019tactical}.\\
\indent Jang first introduced the ANFIS approach by integrating the Fuzzy Inference System (FIS) into the adaptive network architecture ~\cite{jang1993anfis}. An adaptive network is a network consisting of several interconnected nodes linked by directed connections. The outcomes of these flexible nodes are impacted by modifiable elements associated with these nodes. The learning rules determine the appropriate adjustments to these parameters to minimize error. The FIS consists of three main components: a set of rules, a database, and a reasoning process. The rule base comprises fuzzy if-then rules. Take, for instance, a rule like "if the price is low, then the supplier's rating is high," where cheap and high are linguistic variables. The database defines the membership functions used in fuzzy rules, whereas the reasoning mechanism carries out the inference procedure ~\cite{jang1993anfis}.\\
\indent ANFIS merges the benefits of neural networks with fuzzy logic to give a trustworthy tool for decision-making. Fuzzy logic manages uncertainty and imprecision, while neural networks provide robust learning. Multiple areas have utilized this combination to improve prediction and system optimization ~\cite{behnke2020boundary}. ANFIS combines neural network flexibility and fuzzy logic precision for supply chain efficiency, and it also increases the system's ability to foresee and respond to complex and ever-changing supply chain conditions, improving demand prediction, inventory management, and logistics. Due to its flexibility and learning capacity, ANFIS is suitable for unpredictable and fast-changing situations ~\cite{behzadi2018agribusiness}. 
This study utilizes ANFIS to develop an SCM model. The technique comprises the collection of data, the construction of a system model, and the evaluation of performance. ANFIS may lead to higher levels of customer satisfaction, cost savings, and delivery efficiency. Comparing ANFIS against existing SCM systems allows for a clear demonstration of its effectiveness and possible benefits ~\cite{sabri2018exploring}. Essential components of the suggested SCM system include:

•	A blockchain-based governance model is proposed to improve openness and trust within SCM using distributed ledgers and smart contracts. This model aims to reduce fraud and increase financial transparency.

•	The proposed model enables more accurate supply chain forecasts, which will help improve the management of supply and demand uncertainty. This allows for more efficient decision-making compared to state-of-the-art methods.

•	Our model utilizes an ANFIS to maximize economies of scale, lower transportation costs, and enhance inventory control, thus improving overall supply chain performance.
The research is structured to provide a solid framework by addressing current SCM problems and the transformational potential of blockchain and fuzzy neural systems. In the second part, earlier research compiled in a literature review supports a theoretical basis for integrating numerous technologies into SCM. The proposed approach part in section three fully justifies the ANFIS and blockchain-based solutions. Examined is the system's design and implementation; a simulation study is carried out to confirm its efficiency. Section four, Experiment, offers a synthesis of the findings and investigates possible avenues for further investigation. The conclusion will address future activities and bring the present inquiry to a close.
\section{{Literature Review}}\vspace{0.5cm}
In recent years, there has been an increased focus on state-of-the-art technologies to enhance SCM. In this section, we review the literature on SCM and its combination with fuzzy-neural networks and blockchain separately. In the first section, we introduce fuzzy logic and its application in SCM, and then we explain the role of blockchain in supply chain management.\\
blockchain and artificial intelligence in general. In summary, existing reviews that contend specifically with the blockchain and large language model space are not focused on the direction in which they apply the technologies. On the other hand, reviews that concern blockchain and AI as a whole lose the benefits of tighter granularity and focus. 
\subsection{Supply Chain Management and Fuzzy Logic}
\vspace{0.5cm}
Since the 1960s and 1970s, Supply Chain Management (SCM) has seen advancements to enhance the efficiency and quality of organizational processes ~\cite{behzadi2018agribusiness}. The information revolution and evolving consumer expectations during the 1980s and 1990s prompted the formal adoption and improvement of supply chain management techniques in the early 1990s ~\cite{barbosa2018opportunities}.
This discipline tackles the smooth flow of products from the purchase of raw materials (upstream) through manufacture (internal) to the last distribution of items to customers (downstream). Emphasizing the network of several phases, Figure 1 presents the whole supply chain management system view.
    \begin{figure*}[h]
        \centering
        \includegraphics[width=1\linewidth]{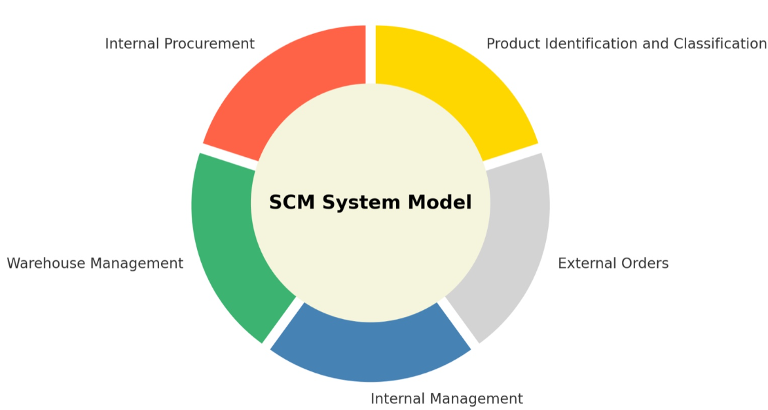}
        \caption{Overview of the supply chain management system.}
        \label{fig:blockchain_components}
    \end{figure*}
Strategic supply chain management is crucial in maximizing operations, reducing costs, and improving service delivery in line with market needs. Fostering better relationships between suppliers, manufacturers, and distributors helps maintain a constant competitive advantage by integrating various supply chain processes ~\cite{behzadi2018agribusiness}, ~\cite{barbosa2018opportunities}. Managing these complex networks is vital in today's global corporate environment, where responsiveness and efficiency are prerequisites to maintaining competitive market positions and ensuring customer satisfaction. The processes of the supply chain management system are depicted in Figure 2.
      \begin{figure}[h]
        \centering
        \includegraphics[width=1\linewidth]{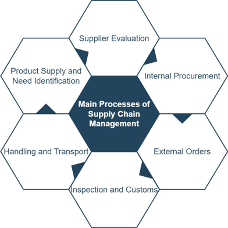}
        \caption{Processes of the Supply Chain Management System.}
        \label{fig:blockchain_components}
    \end{figure}  
 
Blockchain technology was created in 2008 to tackle the issue of digital double-spending in supply chain management ~\cite{treiblmaier2018impact}. It achieves this by offering secure, transparent, and efficient operations. This technology has the potential to revolutionize non-bank financial transactions in the realm of DeFi.
These combined characteristics create a sturdy base for effectively controlling and improving supply chains, helping companies manage the complexity of contemporary markets~~\cite{tonnissen2020analysing}. Blockchain has been applied in various other areas, such as asset management, car manufacturing, and especially in the food industry, to ensure safety and cleanliness~\cite{casino2019systematic, feng2020applying} . Blockchain technology enables transparent, permanent ~\cite{al2019blockchain}, and secure transactions, which enhances traceability and reduces supply chain uncertainty ~\cite{lim2021literature}.\\
\indent For instance, by integrating blockchain with IoT, the halal food supply chain has become more efficient, optimizing ongoing activities and monitoring within the supply chain. This approach ~\cite{azimifard2018selecting} tackles food supply security and reliability issues. Combining the IoT with blockchain is particularly important in healthcare and food, as it improves operational efficiency while ensuring strict compliance with certification criteria.\\
\indent Fuzzy logic connects inputs and outputs through the use of fuzzy inference. This mapping method determines or identifies a specific pattern or substitute as the outcome by applying their membership functions, fuzzy operators, and if-then rules that primarily define fuzzy inference. Two commonly used variations of fuzzy inference systems are the Mamdani technique and the Sugeno method, sometimes known as Takagi Sugeno Kang or TSK ~\cite{de2020blockchain}. These two kinds differ in their approaches to computing the outcomes~\cite{wang2020blockchain}. These systems have proven successful in various disciplines, including but not limited to automobile control, data classification, decision analysis, expert systems, and machine vision. The numerous names used to characterize fuzzy inference systems reflect their wide range of applications. These titles include fuzzy modeling, fuzzy associative memory, fuzzy logic controllers, fuzzy rule-based systems, fuzzy expert systems, and fundamental fuzzy systems. Moreover, neural network modeling has been utilized to predict conditions such as food temperature during transportation, significantly increasing the accuracy and efficiency of supply chain operations ~\cite{sremac2018anfis}. Similarly, artificial neural networks have been employed to optimize warehouse capacities and effectively manage supply chain risks ~\cite{hyndman2018forecasting},~\cite{mercier2018neural}.\\
\indent Implementing fuzzy network analysis and multi-criteria decision-making methods like the Analytic Hierarchy Process (AHP) has further enhanced the logical components of SCM, enabling companies to prioritize crucial elements such as green supplier practices and environmental management ~\cite{varriale2021sustainable, farhadi2023leveraging}.
\subsection{Blockchain in Supply Chain Management}
\vspace{0.5cm}
 \indent In the seafood supply chain, blockchain helps identify instances of adulteration and monitor fish quality, eventually leading to an improvement in traceability from the farm to the customer ~\cite{singh2023revealing}. Models of supply chain finance could help decide how much economic compensation is needed to preserve fertile ground. This blockchain integration with supply chain finance provides a framework for valuing land fairly and precisely by including dynamic elements such as land features and market conditions ~\cite{de2020blockchain}. Blockchain technology guarantees data transparency and integrity, improving theoretical and practical land protection and valuation methods.\\
\indent The transparency and immutability of blockchain transactions help increase stakeholder confidence and responsibility for operational efficiency and regulatory compliance ~\cite{de2020blockchain}. Safe multi-party computing and public-key cryptography help blockchain technology address data privacy concerns ~\cite{singh2023revealing}. \\
\indent By utilizing improved data analysis and decision-making procedures, the digital twin approach which generates a virtual supply chain model—helps maximize real-world performance. Blockchain ensures that the data used in these digital twins is tamper-proof and safe, thereby improving operational efficiency ~\cite{yang2022edge}. Employing real-time data analysis and feedback loops in the physical supply chain increases operational effectiveness and supports large-scale and intricate industry systems ~\cite{wang2020blockchain}. \\
\indent Due to its decentralization, the blockchain records all transactions in an immutable ledger that cannot be changed. This lowers logistics costs and improves efficiency and reliability ~\cite{wang2020blockchain}. This integration simplifies surveillance and decision-making, improving supply chain efficiency ~\cite{nanda2023medical}.\\

\section{Proposed Method}
\vspace{0.5cm}

\indent Managing the logistics of transporting products and services across international marketplaces presents several substantial difficulties in today's interconnected world. These challenges include volatile demand, intricate logistical barriers, and unanticipated interruptions, such as those induced by the COVID-19 epidemic. Blockchain technology, the Adaptive Neuro-Fuzzy Inference System (ANFIS), and the Internet of Things (IoT) are proposed solutions for constructing a robust, transparent, and efficient supply chain management system.\\
\indent Effectively controlling complexity and supply chain management (SCM) hazards depends on incorporating ANFIS with data that comes from IoT into the blockchain paradigm. As shown in Figure 3, the integrating system begins with real-time data collection from IoT devices deployed throughout the supply chain, including transaction records, inventory levels, delivery schedules, costs, and so on, as required by the system via sensing and collecting data through various means—vision, text, and other formats—based on the model's specific needs and acceptable inputs.\\

      \begin{figure}[h]
        \centering
        \includegraphics[width=1\linewidth]{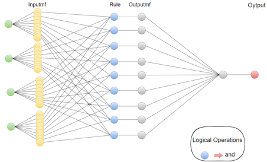}
        \caption{Architecture of the integrated system}
        \label{fig:fig3}
    \end{figure}  
\indent In the second step, the raw data may contain noise and irrelevant information, which will be preprocessed for cleaning and normalization data; the fuzzy C-means (FCM) method is employed to cluster data with diverse memberships ~\cite{krinidis2010robust}. \\
\indent The third step, the ANFIS model, provides a predictive framework for the fundamental elements of supply chain management—demand prediction, inventory control, and logistics optimization. While analyzing current data, the system continuously acquires new data; ANFIS processes this information by combining the human-like reasoning of fuzzy logic with the learning ability of neural networks, thereby enhancing its capacity to deliver accurate predictions and informed decision-making over time ~\cite{sremac2018anfis}.\\
\indent During step four, blockchain technology considers the ANFIS output an equal opinion among the SCM stakeholders, promoting a cooperative decision-making process. This method guarantees that all essential participants are not just engaged but actively assessing and reaching a consensus on choices prior to execution. The level of confidence in distributed supply chain management relies heavily on the transparency, security, and traceability of blockchain technology. This ensures that every stakeholder is accountable and actively involved in the decision-making process.\\
\indent However, at this stage, smart contract demonstrates their effectiveness by autonomously carrying out procedures according to ANFIS determinations.
 For instance, a smart contract may independently initiate a purchase order with suppliers if the ANFIS model forecasts greater inventory needs. This not only ensures the speed of the process but also the system's accuracy, which guarantees the safe, open, and quick management of financial transactions. To monitor the reliability of these exchanges, the system also combines a distributed finance (DeFi) mechanism. The system uses a multi-signature technique that requires several authorizations for every transaction, minimizing the potential for fraud and promoting confidence among all parties involved ~\cite{treiblmaier2018impact}.
 \subsection{Fuzzy Logic Functions and Clustering Techniques}
 \vspace{0.5cm}

\indent In order to properly implement the ANFIS model, certain fuzzy logic functions and clustering algorithms are used. The used functions consist of Gaussian and triangular. When only using fuzzy logic, triangular functions are implemented. However, when integrating the neuro-fuzzy section, a mix of triangular and Gaussian functions is utilized. The programming strategy used in this study refers to a fuzzy inference system known as FCM, which utilizes the C-means clustering algorithm inside ANFIS loops ~\cite{sremac2018anfis, hyndman2018forecasting}. The main distinction between classical and fuzzy clustering lies in the fact that with fuzzy clustering, a sample may be assigned to many groups ~\cite{mercier2018neural}.\\
\indent On the other hand, in the classical C-means algorithm, the number of clusters c is predetermined in this algorithm Krinidis and Chatzis ~\cite{krinidis2010robust}. The objective function defined for this algorithm is as follows:\\
\begin{equation}
  J = \sum_{i=1}^{c} \sum_{k=1}^{n} u_{ik}^m d_{ik}^2 = \sum_{i=1}^{c} \sum_{k=1}^{n} u_{ik}^m \| x_k - v_i \|^2    \label{eq1} 
\end{equation}

\indent In this equation, m is a natural number greater than one, often chosen as two. $u_{ik}$ represents the degree of membership of sample k in the cluster i, and $v_i$ represents the cluster center i. The term $‖x_v-v_i ‖$  indicates the similarity (distance) between the sample and the cluster center, which can be measured using any function that indicates the similarity between the sample and the cluster center.the matrix $U_{ik}$ can be defined with c rows and n columns, where the components can take any value between zero and one ~\cite{hyndman2018forecasting}.  If all components of matrix U are either zero or one, the algorithm behaves like classical C-means. However, the components of matrix U can take any value between zero and one, with the sum of the elements of each column required to equal 1:\\
 \begin{equation}
  \sum_{i=1}^{c} u_{ik} = 1, \quad \forall k = 1, \ldots, n \label{eq2}  
 \end{equation}

\indent This condition means that the total membership of each sample in C clusters must equal one. Using the above condition and minimizing the objective function, the following relationships exist:

\begin{equation}
 u_{ik} = \left( \sum_{j=1}^{c} \left( \frac{\| x_k - v_i \|}{\| x_k - v_j \|} \right)^{\frac{2}{m-1}} \right)^{-1} \label{eq3}   
\end{equation}

\begin{equation}
v_i = \frac{\sum_{k=1}^{n} u_{ik}^m x_k}{\sum_{k=1}^{n} u_{ik}^m}  \label{eq4}  
\end{equation}

\indent the advantages of GENEFIS3 are considered significant. One of these advantages of fuzzy clustering is that data can belong to more than one cluster with different membership degrees. GENEFIS3 performs its functions by extracting a series of rules that model the behavior of the data. The rule extraction method initially uses the FCM function, which is used to determine the number of fuzzy sets and membership functions for all previous sections. The number of clusters is collected by a comprehensive search method, and the forecast for the previous period is$ x_t$. This forecast is simply a weighted set between the last observation $F_t$, where $0<\alpha <1$ which referred to as stabilization:\\
\begin{equation}
  F_{t+1}=\alpha x_t (1-\alpha ) F_t  \label{eq5}
\end{equation}
           
\indent Due to the existence of recursive relationships between $F_(t)$and $F_{t+1}$, $F_{t+1}$ can also be displayed in another form, such as the following relationship. In expressing the relationship, exponential smoothing assigns the most weight to $x_t$ and less to previous observations. Additionally, this relationship does not require keeping data before period t to estimate the demand for the next period, making it a simple method:
\begin{equation}
 F_{t+1}=\alpha x_t+\alpha (1-\alpha ) x_{t-1}+\alpha (1-\alpha )^2 x_{t-2}+ ..  \label{eq6}  
\end{equation}
                                   
All that is needed is $x_t$ and the previous forecast, $F_t$. The exponential smoothing relationship can be expressed similarly to the following relationship:
\begin{equation}
F_{t+1} = F_t + \alpha (x_t - F_t)  \label{eq7}
\end{equation}
   
\indent This relationship shows that the forecast for period $t+1$ equals the sum of the forecast for the previous period t and the product of the forecast error in period t by a discount factor $\alpha$ . In the next section, we separately estimate the time distance between previous transactions Pt and the size of each transaction $Z_t$. If no demand has occurred in the review period t, then at the end of period t, the estimates $p_t$ and $z_t$ do not change. If a transaction occurs, $z_t \neq 0$, he estimates are updated according to the following relationship:\\
\begin{equation}
 \hat{z}_t = \alpha z_t + (1-\alpha) \hat{z}_{t-1}
  \label{eq8}
 \hat{p}_t = ap_t + (1-a) \hat{p}_{t-1}, \quad 0 \le a \le 1                    
\end{equation}

According to equation (8), $\alpha$  is the smoothing constant. Therefore, the demand forecast for each period at time t for $c_t$ Is calculated as follows:
\begin{equation}
  c_t = \frac{\hat{z}_t}{p_t}      \label{eq9} 
\end{equation}
Finally, using the ANFIS approach, the demand forecast for each period at time t for  $s_r$ is calculated as follows:
\begin{equation}
 s_t = \frac{(1 - \frac{alpha}{2}) \hat{z}_t}{\hat{p}_t - \frac{alpha}{2}} \label{eq10}
\end{equation}
\begin{figure}[h]
        \centering
        \includegraphics[width=1\linewidth]{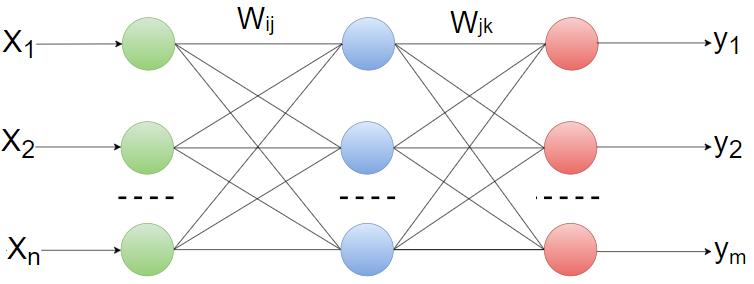}
        \caption{Proposed ANFIS model for supply chain management}
        \label{fig:fig4}
    \end{figure}
The model presented in this research has an ANFIS structure as shown in Figure \ref{fig:fig4}.
In Figure 4, the inputs are denoted as $x_n$, where n represents the number of inputs, which will pass through the input layer. From there, they pass through a hidden layer, where weight vectors, represented as $w_ij$, are applied to the input vector. The weight vectors from the hidden layer to the outputs $(y_n)$ are represented with $w_jk$ . These weight vector parameters define the relationships in a fuzzy manner. This process involves training and defuzzification to transform fuzzy rules into clear and actionable outputs.
The most effective way to attain the best economic results in many market scenarios for all variables like sales income, cost structures, and marketing expenses is the O-game strategy. Equations 11 and 12 help to show almost all the strategic financial operations of the supplier and retailer in the supply chain
\begin{equation}
\pi_s^0 = \max_{A^0} \left\{ Q^0 (e - c_s) - \frac{ca(A^0)^2}{2} - (\lambda_s Q^0 + A^0) c_T \right\}   \label{eq11}
\end{equation}
\begin{multline}\label{eq:21} 
\pi_R^0 = \max_{P^0, Q^0} \{ [aQ^0 (1-\lambda) - \beta p^0 + \theta A^0 (1-\phi)](p^0 - c_R) \\ -  Q^0 (w + \lambda_R c_T) - \frac{c}{(Q^0)^2} \}
\end{multline}
In the financial model of the supply chain, supplier revenue $w_cs$ is calculated by subtracting the production cost $c_s$ from total revenue with additional expenses incurred from online service investments $c_A$. These investments are modeled to have a quadratic relationship reflecting their concave impact on company performance, moderated by transaction volume $Q^0$ and online effort $A^0$, with a scaling parameter $\lambda_s$ to adjust economic impacts ~\cite{mercier2018neural}. \\
\indent This interface streamlines IoT and DeFi supply chain interactions. DeFi uses multiple signatures to provide security since transactions to be carried out need approvals from several private keys. Shared ledger registration and validation also become easier.\\
\subsection{The Blockchain Structure and Integration with ANFIS}
 \vspace{0.5cm}

\indent The implementation of blockchain technology has led to substantial changes in the structure and management of businesses, since it enables decentralization and automation. Nevertheless, the integration of blockchain technology does entail specific hazards. The risks encompass vulnerabilities in the consensus mechanisms, software governance, and critical administration, which have the potential to compromise the integrity of data and introduce potential points of failure. This highlights the importance of implementing intrusion detection systems, which offer a cautious and methodical method for incorporating technology. \\
Figure 5 illustrates a detailed model showcasing the potential utilization of blockchain technologies in a chicken supply chain.
\begin{figure*}[h]
        \centering
        \includegraphics[width=1\linewidth]{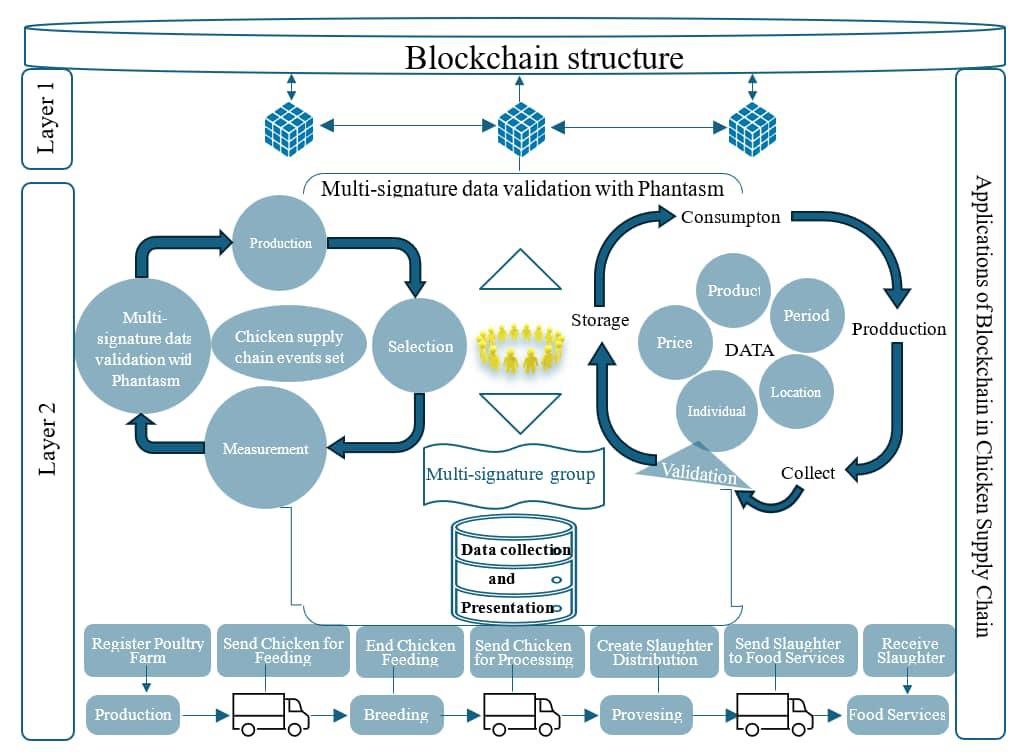}
        \caption{Blockchain structure with a multi signature protocol}
        \label{fig:fig5}
    \end{figure*}
\indent Figure 5 exhibits a blockchain topology that securely underpins a chicken supply chain operation by offering all involved a trustworthy and tamper-proof record. It can analyze intricate data and generate precise predictions, which subsequently inform supply chain operations to guarantee the unchangeable recording of every transaction and data input, thus creating an inviolable chronicle on which all those engaged may rely.  In Layer 1 of Figure 5, ANFIS, a powerful decision making model, validates the results by thoroughly analyzing each decision made by the system's stakeholders. This approach reduces computational complexity, making it easier for operators to manage complex tasks, and enhances market control through the efficient use of the DeFi blockchain solution ~\cite{farhadi2024domain}. Data is only stored on the blockchain after the necessary signatures are obtained using the Phantom technique in Layer 2. This process helps reduce errors in critical management, software control, and consensus mechanisms, improving decision-making accuracy. 
On the right side of Layer 2, in Figure 2, there is a continuous monitoring structure for validating and updating all pertinent data by collecting important information about production type, period, location, price, and individual validation processes. Moreover, integrating every stage of the supply chain with a responsible person helps the operators engaged in the process—farm managers, processors, or distributors—monitor themselves. \\
\indent The left side of the second layer shows the continuous flow of important events across the chicken supply chain to generate basic statistics on the chickens and determine which chicks should continue through the process. Additionally, the measurement phase captures and evaluates health, feed quality, and growth rates to guide future actions. At the core of this cycle is the Chicken Supply Chain Events Set, which ensures that every activity within the supply chain is documented.
The lower portion of Figure 5 depicts the logistical process of chicken production throughout the supply chain, from poultry farm registration to final delivery to food services. Comprehensive documentation at every step provides crucial information for tracking and making decisions. 
Figure 6, a supply chain model, comprises four basic nodes—supplier, producer, distributor, and retailer. The model incorporates particular product data. The values of these qualities are interconnected and exhibit a strong correlation. Other qualities include transportation capacity. Until the values are in the hands of the retailer at about 90\% and in other points of the supply chain at equal levels, they are constantly changing since they are distributed throughout the supply chain. The data are taken from sources ~\cite{farhadi2024domain} and ~\cite{farhadi2023unsupervised} collected via field study by examining their outputs and parameters. The proposed supply chain model is outlined in Figure 6 ~\cite{mahalegi2024generative}.
\begin{figure}[h]
        \centering
        \includegraphics[width=1\linewidth]{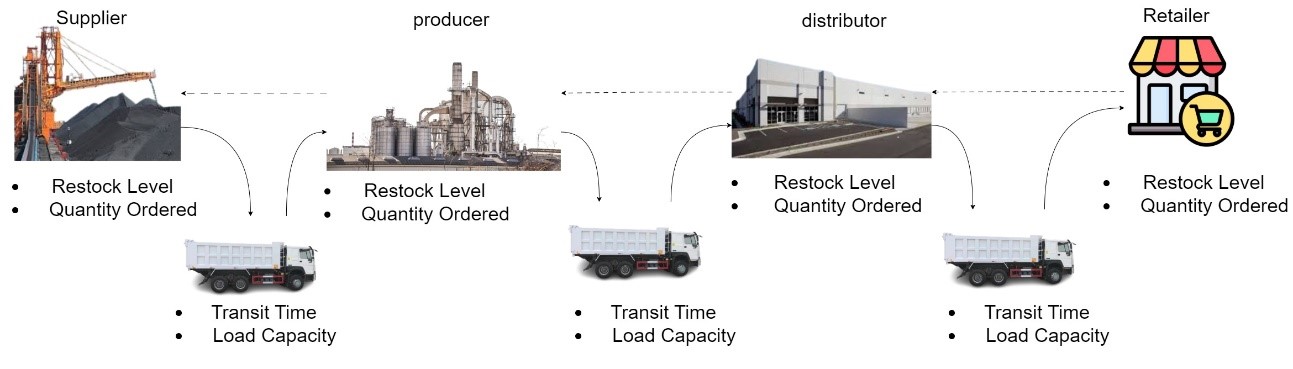}
        \caption{Supply chain model}
        \label{fig:fig6}
    \end{figure}
\section{Experiment}
\vspace{0.5cm}

\indent The preceding section presented a novel paradigm for managing supply chains inside a framework based on blockchain technology. Continuing with the previous information, this section describes a simulation carried out using MATLAB software. The main aim of this section is to discuss the results obtained from the simulation. The suggested solution utilizes a neuro-fuzzy method to model supply chain management, and the results are extensively evaluated. The decision to choose a neuro-fuzzy network instead of other recognized techniques like the Hidden Markov Model or game theory methods is based on the inherent uncertainty that exists in the supply chain model. It is essential to undergo thorough data training, a process that we have meticulously followed, in order to obtain more precise and dependable outcomes, hence increasing confidence in the effectiveness of our methodology.\\
\indent The input layer can only include inventory and order receipt until arrival or extend the time range in temporal data. In this case, the input layer is configured with all inventory quantities. The structure of the neuro-fuzzy network is shown in Figure 4 of the previous section. It is assumed that nine inputs range from $x_1$ to $x_9$.
\begin{table*}[t]
\centering
\caption{Input Data for Inventory Management Over 30 Days}
\begin{tabular}{|l|c|c|c|c|c|c|c|c|c|}
\hline
\textbf{Data} & \textbf{1} & \textbf{2} & \textbf{3} & \textbf{4} & \textbf{5} & \textbf{6} & \textbf{7} & \textbf{8} & \textbf{9} \\ \hline
Quality, P1 (\%) & 96.82 & 100.0 & 96.46 & 96.77 & 96.22 & 96.79 & 98.86 & 97.62 & 100 \\ \hline
Material, P2 (\%) & -9.63 & -10.24 & -9.59 & -9.63 & -9.57 & -9.63 & -9.83 & -9.71 & -10.32 \\ \hline
Logistic, P3 (\%) & 10.89 & 11.59 & 10.86 & 10.89 & 10.93 & 10.99 & 11.13 & 10.99 & 11.67 \\ \hline
Purchase completed, P4 (\%) & 97.3 & 100 & 96.95 & 97.26 & 96.71 & 97.28 & 99.35 & 98.11 & 100 \\ \hline
Order plan, L1(\%) & 90.69 & 96.41 & 90.35 & 90.64 & 90.13 & 90.66 & 92.61 & 91.41 & 97.14 \\ \hline
Order speed, L2 (day) & 3.21 & 3.43 & 3.20 & 3.21 & 3.19 & 3.21 & 3.28 & 3.41 & 3.44 \\ \hline
Delivery on time, C1 (\%) & 95.46 & 100 & 95.1 & 95.41 & 94.48 & 95.43 & 97.47 & 96.25 & 100 \\ \hline
Volume, C2 (\%) & 92.63 & 98.8 & 92.29 & 92.59 & 92.06 & 92.61 & 94.59 & 93.4 & 99.23 \\ \hline
Node satisfaction, C4 (\%) & 3.4 & 3.62 & 3.39 & 3.4 & 3.38 & 3.4 & 3.48 & 3.43 & 3.64 \\ \hline
Outbound error, S1 (\%) & 4.69 & 5.28 & 4.46 & 4.96 & 4.93 & 4.96 & 5.07 & 5.00 & 5.32 \\ \hline
Damage rate, S2 (\%) & 0.01 & 0.01 & 0.01 & 0.01 & 0.01 & 0.01 & 0.01 & 0.01 & 0.01 \\ \hline
Turnover times, S3 (times) & 1.20 & 1.24 & 1.19 & 1.20 & 1.19 & 1.20 & 1.22 & 1.21 & 1.28 \\ \hline
Zero inventory (times) & 0.51 & 0.53 & 0.51 & 0.51 & 0.51 & 0.52 & 0.51 & 0.51 & 0.54 \\ \hline
\end{tabular}
\label{table1}
\end{table*}
\indent In Table 1, a series of key features of supply chain data that has been utilized is considered, including quality, goods, logistics, complete procurement, order scheduling, order speed, on-time delivery, goods value, delivery error, delivery satisfaction, output error, damage rate, goods in circulation, and zero inventory , the table columns numbered 1 to 9 reflect several phases in the supply chain. Covering actions across stores, distributors, manufacturers, and suppliers. These phases match the several nodes as the main inputs and outputs of the Internet of Things (IoT), where data is gathered and monitored in different time interval, which is shown in Figure 7.
\begin{figure*}[h]
        \centering
        \includegraphics[width=1\linewidth]{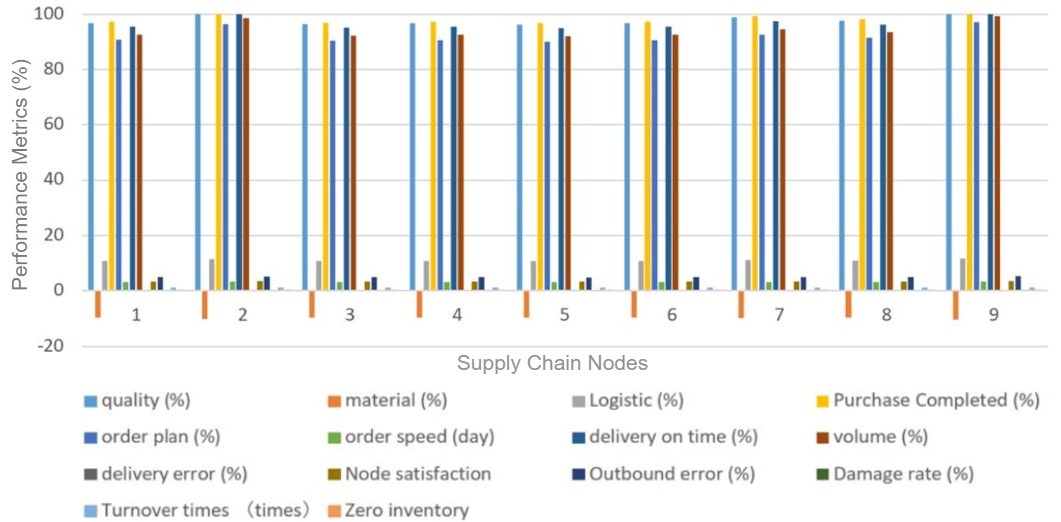}
        \caption{The input data}
        \label{fig:fig7}
    \end{figure*}
\indent The bar charts in Figure 7 graphically show several supply chain measures including quality, logistics, completed purchases, order planning, order speed, on-time delivery, product value, delivery errors, satisfaction levels, production errors, damage rate, inventory circulation, and zero inventory. Every measure has a color coding for clarity. Table 2 also provides comprehensive neuro-fuzzy network parameter information.\\
\indent Based on Table 2 and the chart in Figure 7, the number of inputs is nine, and the number of outputs is six, which indicates the level of satisfaction in supply chain management, categorized as average, sound, and excellent. the y-axis representing the percentage during the data processing time for supply chain management using the neuro-fuzzy network in the blockchain platform with the DeFi multi-signature protocol; the y-axis represents the different time intervals in ten phases within the supply chain. However, during simulation with the neuro-fuzzy network, the inputs are considered eight, excluding the zero inventory (SC). zero inventory is not considered an input but a metric for assessing the total inventory in the warehouse. The total number of items is around six thousand.
 \begin{table*}[t]
    \centering
    \caption{parameters of neural-fuzzy network}
    \label{tab:anfis-config}
    \begin{tabular}{@{}ll@{}}
        \toprule
        \textbf{Parameter} & \textbf{Value} \\ \midrule
        ANFIS Training Data Count & Twelve data sets (nine used) \\
        ANFIS Test Data Count & Eight data sets (four usable) \\
        ANFIS Output Count & Six items \\
        Fuzzy Linguistic Terms Sampling Count & 20 \\
        Initial ANFIS Accuracy in Percentage & 0.001 \\
        ANFIS Threshold for Training and Testing & 1.3 \\
        ANFIS Alpha Value & 0.1 \\
        ANFIS Epoch Threshold Value & 0.6 \\ \bottomrule
    \end{tabular}
\end{table*}
\indent When data is applied as input and the supply chain system is modeled using a neuro-fuzzy network, the initial results of supply chain management optimization on the blockchain platform can be observed in Figure 8. This optimization utilizes the DeFi multi-signature protocol within the Internet of Things network, as demonstrated by the proposed approach. 
\begin{figure}[h]
        \centering
        \includegraphics[width=1\linewidth]{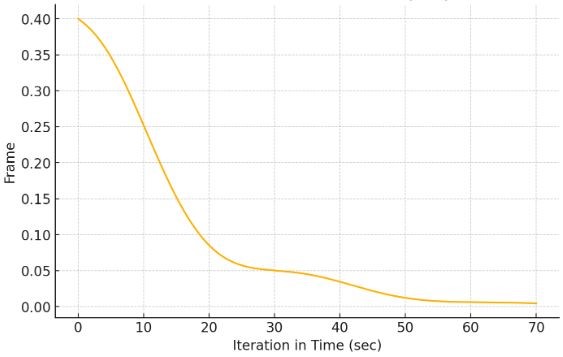}
        \caption{The optimal rate  of the SCM system}
        \label{fig:fig8}
    \end{figure}
Figure 8, is the exact output generated by the model. Similarly, the outputs for predicting each influential factor in the data, according to Table 1, are presented in Figures 9 to 14
\begin{figure}   \centering
    \begin{minipage}{0.3\textwidth}
    \frame{\includegraphics[width=1\textwidth]{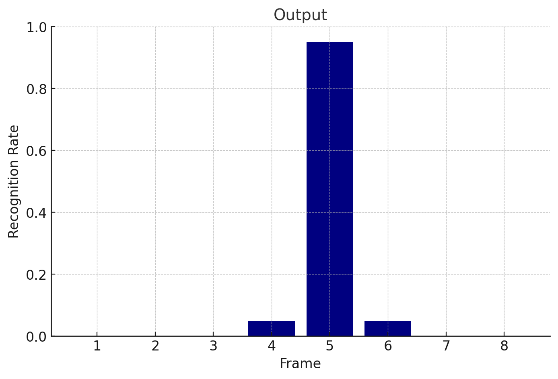}}
    \caption{The first output}
    \label{fig:fig3_1}
    \end{minipage}
    \begin {minipage}{0.3\textwidth}
     \frame{\includegraphics[width=1\textwidth]{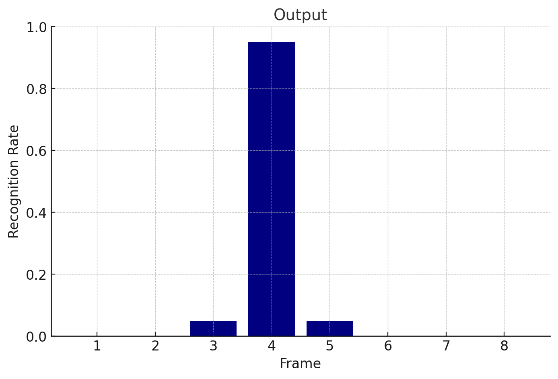}}
       \caption {The second output}
    \label{fig:fig3_2}
   \end{minipage}
\end{figure}
 \begin{figure}   \centering
    \begin{minipage}{0.3\textwidth}
    \frame{\includegraphics[width=1\textwidth]{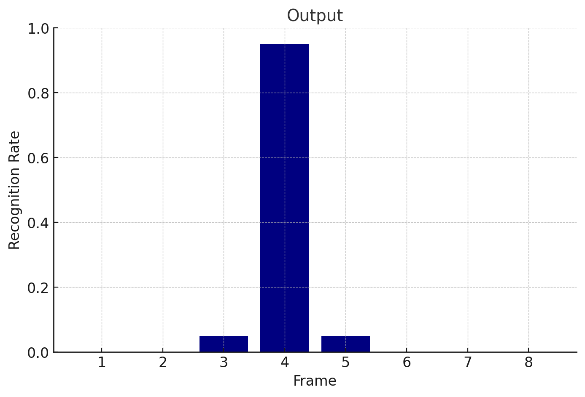}}
    \caption{The third output}
    \label{fig:fig3_3}
    \end{minipage}
    \begin {minipage}{0.3\textwidth}
     \frame{\includegraphics[width=1\textwidth]{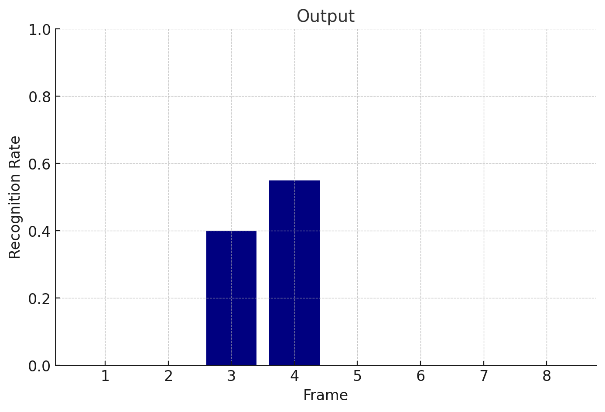}}
       \caption {The forth output}
    \label{fig:fig3_4}
   \end{minipage}
\end{figure}
   \begin{figure}   \centering
    \begin{minipage}{0.3\textwidth}
    \frame{\includegraphics[width=1\textwidth]{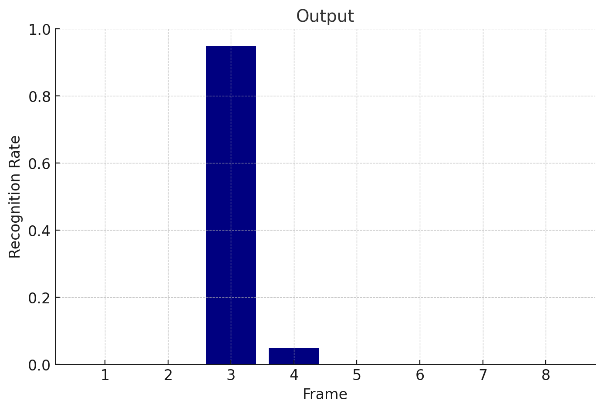}}
    \caption{The fifth output}
    \label{fig:fig3_5}
    \end{minipage}
    \begin {minipage}{0.3\textwidth}
     \frame{\includegraphics[width=1\textwidth]{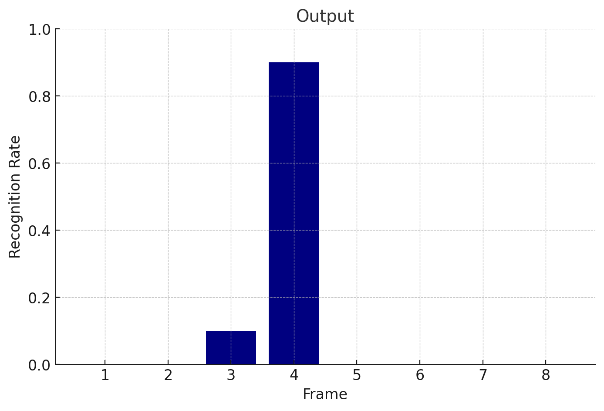}}
       \caption {The sixth output}
    \label{fig:fig3_6}
   \end{minipage}
\end{figure}
The outputs are complete up to this point, as the system had determined six outputs . However, the subsequent outputs for the intended inputs are also shown in Figures 15 and 16.
  \begin{figure}   \centering
    \begin{minipage}{0.3\textwidth}
    \frame{\includegraphics[width=1\textwidth]{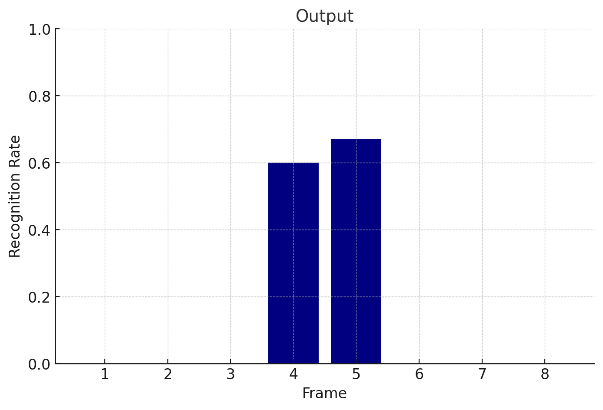}}
    \caption{The seventh output}
    \label{fig:fig3_7}
    \end{minipage}
    \begin {minipage}{0.3\textwidth}
     \frame{\includegraphics[width=1\textwidth]{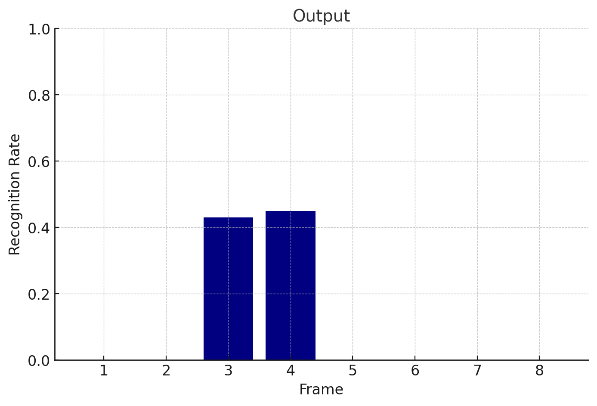}}
       \caption {The eight output}
    \label{fig:fig3_8}
   \end{minipage}
 
\end{figure}
The results are shown in Figure 17 during processing with the neuro-fuzzy network.
\begin{figure}[h!]
        \centering
        \includegraphics[width=1\linewidth]{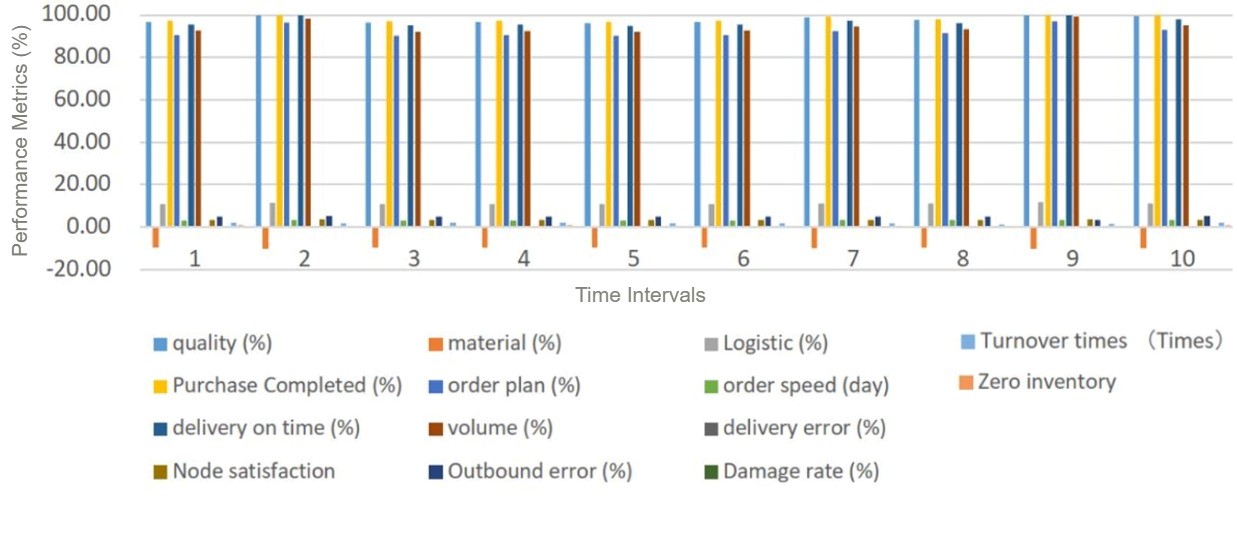}
        \caption{The data processing time for the DeFi multi-signature  protocol}
        \label{fig:fig7}
    \end{figure}
 Table 3 lists the numerical results and statistical analysis of Figure 17 for recording and analysis. 
 \begin{table*}[t]
\centering
\caption{The output of the DeFi multi-signature protocol}
\begin{tabular}{|l|c|c|c|c|c|c|}
\hline
\textbf{Indicator} & \textbf{Data one} & \textbf{Data two} & \textbf{Data three} & \textbf{Data four} & \textbf{Data five} & \textbf{Data six} \\ \hline
Quality, P1(\%) & 95.07 & 97.34 & 95.52 & 99.50 & 100 & 92.65 \\ \hline
Material, P2(\%) & 6.46 & -0.21 & 3.20 & -9.90 & -10.32 & -1.19 \\ \hline
Logistic, P3 (\%) & 12.86 & 12.37 & 11.69 & 11.20 & 11.67 & 11.34 \\ \hline
Completed Purchase, P4(\%) & 84.86 & 94.87 & 99.06 & 100 & 100 & 96.11 \\ \hline
Order Planning, L1(\%) & 87.20 & 100 & 98.10 & 93.20 & 97.14 & 95.15 \\ \hline
Order Speed, L2 (\%) & 6.64 & 4.93 & 4.96 & 4.30 & 3.44 & 3.52 \\ \hline
Delivery on Time, C1 (\%) & 92.00 & 100 & 96.61 & 100 & 100 & 93.71 \\ \hline
Volume, C2 (\%) & 96.70 & 96.55 & 90.21 & 95.61 & 99.23 & 93.32 \\ \hline
Delivery Error, C3(\%) & 0.20 & 5.12 & 0 & 0 & 0 & 3.86 \\ \hline
Node satisfaction, C4(\%) & 3.64 & 3.27 & 3.27 & 3.50 & 3.64 & 3.37 \\ \hline
Output Error, S1 & 3.58 & 4.72 & 2.58 & 3.10 & 3.32 & 3.50 \\ \hline
Damage Rate, S2(\%) & 0.53 & 0.16 & 0.07 & 0.01 & 0.01 & 0.22 \\ \hline
Loads in Circulation, S3(times) & 0.92 & 1.64 & 1.64 & 2.23 & 1.58 & 1.65 \\ \hline
Zero inventory, S4(times) & 0.30 & 0.39 & 0.50 & 0.51 & 0.48 & 0.50 \\ \hline
Supply Management Assessment & Medium & Good & Good & Perfect & Perfect & Good \\ \hline
Efficiency & (0,0,1,0) & (0,1,0,0) & (0,0,1,0) & (1,0,0,0) & (1,0,0,0) & (0,1,0,0) \\ \hline
\end{tabular}
\end{table*}
\indent After implementing the proposed model and methodology presented in section three, the results from Table 3 reveal two significant issues pertinent to these tables, which pertain to evaluating supply chain management alongside efficiency. The binary representation of efficiency in Table 3 shows whether specified efficiency criteria have been met, with binary (0 or 1) corresponding to a criterion. The configuration (0, 0, 0, 1), which is the most efficient state according to references ~\cite{mercier2018neural} and ~\cite{varriale2021sustainable} in Columns 4 and 5 of the table contain this configuration, which results from a thorough evaluation. It shows that all required performance criteria have been satisfied, strengthening the supply chain management assessment. by evaluating quality, material efficiency, and delivery accuracy—essential performance parameters. 
The output data produced by our model is pertinent to the chicken distribution industry, as it precisely depicts the ever-changing nature of the supply chain in real-time. Nevertheless, the underlying ideas of the neuro-fuzzy model are not limited to this field. These insights can be applied to other businesses, helping to improve supply chain efficiency in many sectors. 
 \begin{figure}[h!]
        \centering
        \includegraphics[width=1\linewidth]{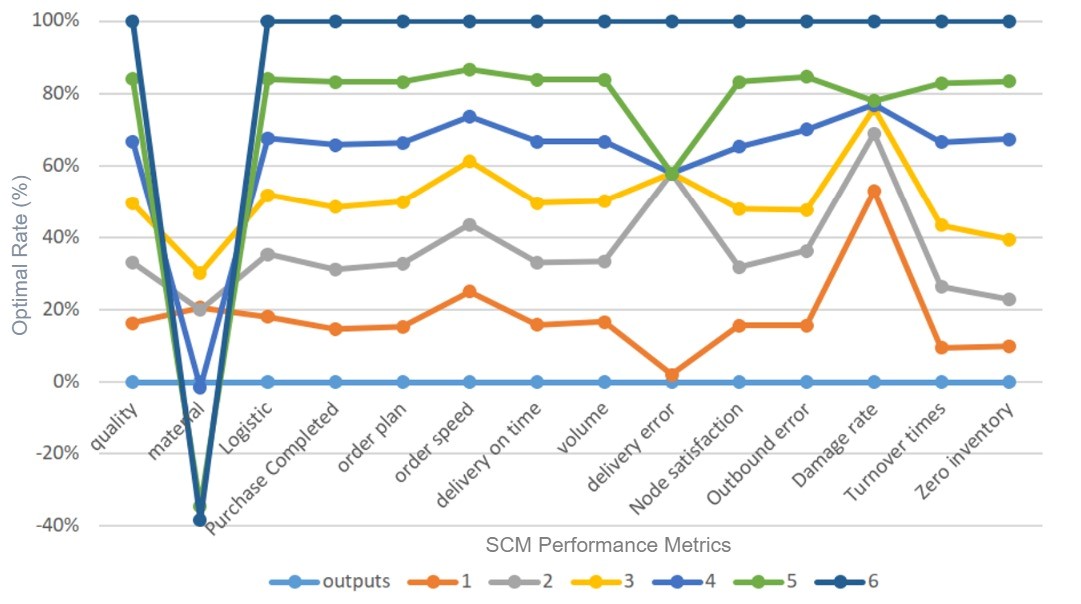}
        \caption{output of supply chain management with the suggested approach}
        \label{fig:fig18}
    \end{figure}
Ultimately, transactions within supply chain management must undergo a realistic assessment. Figure 19 illustrates the system transaction results from the proposed methodology. The x-axis denotes transaction values, while the y-axis indicates the degree of fuzziness, which varies between 0 and 1.   
\begin{figure}[h!]
        \centering
        \includegraphics[width=1\linewidth]{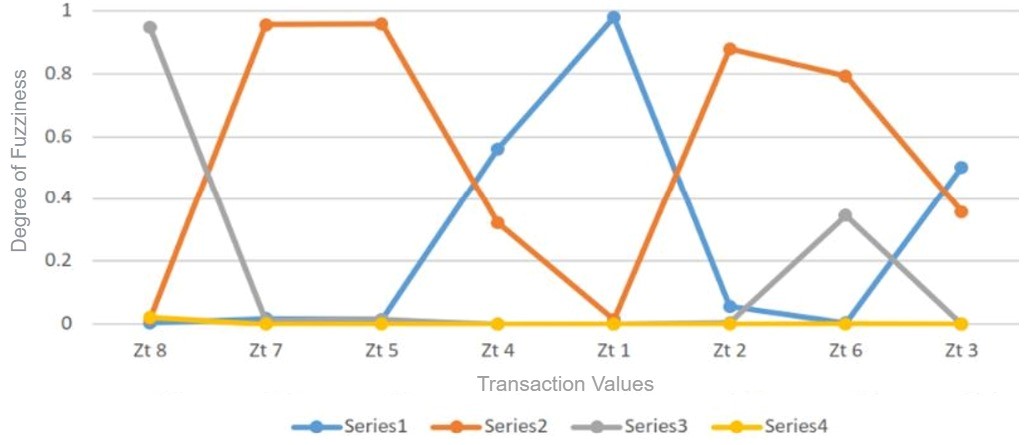}
        \caption{The transaction results from the proposed methodology }
        \label{fig:fig19}
    \end{figure}
From the acquired findings in Table 4, it can be inferred that the input data has been simulated and optimized in unfamiliar circumstances in 8 states $Z_t$ in 4 different metrics, resulting in enhancements in certain aspects while displaying deteriorations in others. The investigation demonstrates that incorporating the neural-fuzzy network into a blockchain platform and implementing a multi-signature DeFi protocol with a smart contract structure greatly improves the optimization capabilities of supply chain management. $Z_t$ 8 stands out as a state that excels in various essential areas such as quality, material efficiency, logistics, and purchase completion. This makes it an excellent example to measure the effectiveness and balance of supply chain activities.  
\begin{table*}[t]
\centering
\caption{Transactions in supply chain management  with the proposed approach}
\begin{tabular}{|l|c|c|c|c|}
\hline
\textbf{State} & \textbf{Metric 1} & \textbf{Metric 2} & \textbf{Metric 3} & \textbf{Metric 4} \\ \hline
Zt 8 & 0.0048 & 0.185 & 0.9484 & 0.0218 \\ \hline
Zt 7 & 0.0172 & 0.9465 & 0.0118 & 0.0001 \\ \hline
Zt 5 & 0.0147 & 0.9598 & 0.0145 & 0.0001 \\ \hline
Zt 4 & 0.5625 & 0.3227 & 0.0003 & 0.0 \\ \hline
Zt 1 & 0.981 & 0.0145 & 0.0 & 0.0 \\ \hline
Zt 2 & 0.0039 & 0.7050 & 0.3461 & 0.0011 \\ \hline
Zt 3 & 0.5034 & 0.359 & 0.0004 & 0.0 \\ \hline
Zt6 & 0.0039 & 0.7036 & 0.0042 & 0.0001 \\ \hline
\end{tabular}
\end{table*}

Additionally, the proposed work is compared with the reference ~\cite{farhadi2023leveraging}, provides a detailed quantitative examination of how blockchain affects supply chain indicators .

\begin{table*}[h]
\centering
\caption{The comparison results of the proposed method with reference ~\cite{farhadi2023leveraging}}
\scriptsize
\begin{tabular}{| c|c|c|c|c|}
\hline
\multirow{3}{*}{\textbf{Parameter}} & \multicolumn{2}{c|}{\textbf{Value}} & \multicolumn{2}{c|}{\textbf{Transaction Time as Percentage}} \\ \cline{2-5}
 & \multirow{2}{*}{\textbf{Proposed Method}} & \multirow{2}{*}{\textbf{~\cite{farhadi2023leveraging} Reference}} & \textbf{Proposed Method} & \textbf{~\cite{farhadi2023leveraging} Reference} \\ 
 &  &  &  &  \\ \hline
Average delivery time to retailers & 37 Minute & 40 Minute & 3.8\% & 4.1\% \\ \hline
\multirow{2}{*}{Reorder Time} & \multirow{2}{*}{4 Days} & \multirow{2}{*}{Five days} & \multirow{2}{*}{3.5\%} & \multirow{2}{*}{3.7\%} \\
 &  &  &  &  \\ \hline
\multirow{2}{*}{Number of products requested per order (for more products)} & \multirow{2}{*}{Eleven products} & \multirow{2}{*}{Six products} & \multirow{2}{*}{8.74\%} & \multirow{2}{*}{9.8\%} \\
 &  &  &  &  \\ \hline
\end{tabular}
\end{table*}

\section{Conclusion}
\vspace{0.5cm}
This paper investigated the improvement of supply chain management (SCM) using blockchain technology, fuzzy logic, and neural networks inside an Internet of Things (IoT) architecture. Simulations showed how the Adaptive Neuro-Fuzzy Inference System (ANFIS) may be included in a blockchain platform using a DeFi multi-signature protocol, especially in Iran's meat sector. This integration sought to improve supply chain data-sharing capacity, scalability, and openness.\\
\indent The findings demonstrated considerable improvements in crucial supply chain parameters, indicating a promising future. Specifically, the average delivery time to merchants fell by 7.5\%, the interval between reorders declined by 20\%, and the number of products requested per order increased by 45.5\%. The data illustrate the capability of the suggested approach in optimizing supply chain procedures.
Nevertheless, the study recognizes that its conclusions are subject to limitations stemming from assumptions about optimal circumstances and restrictions in available data, thereby impacting the broader relevance of the results. Future research should prioritize tackling the obstacles associated with the widespread implementation of blockchain and ANFIS in diverse businesses. Moreover, it is essential to do additional research on sophisticated machine learning techniques and implement the neuro-fuzzy model in various industries to assess its overall influence on the efficiency and efficacy of supply chain operations.
The integration has been proven to improve operations' efficiency and efficacy, creating a strong basis for future developments in supply chain management by utilizing these emerging technologies

%
\printbibliography  

@article{nanda2023medical,
  title={Medical supply chain integrated with blockchain and IoT to track the logistics of medical products},
  author={Nanda, Saroj Kumar and Panda, Sandeep Kumar and Dash, Madhabananda},
  journal={Multimedia Tools and Applications},
  volume={82},
  number={21},
  pages={32917--32939},
  year={2023},
  publisher={Springer}
}

@article{sabri2018exploring,
  title={Exploring the impact of innovation implementation on supply chain configuration},
  author={Sabri, Yasmine and Micheli, Guido JL and Nuur, Cali},
  journal={Journal of Engineering and Technology Management},
  volume={49},
  pages={60--75},
  year={2018},
  publisher={Elsevier}
}

@article{oh2019tactical,
  title={Tactical supply planning in smart manufacturing supply chain},
  author={Oh, Jisoo and Jeong, Bongju},
  journal={Robotics and Computer-Integrated Manufacturing},
  volume={55},
  pages={217--233},
  year={2019},
  publisher={Elsevier}
}

@article{jang1993anfis,
  title={ANFIS: adaptive-network-based fuzzy inference system},
  author={Jang, J-SR},
  journal={IEEE transactions on systems, man, and cybernetics},
  volume={23},
  number={3},
  pages={665--685},
  year={1993},
  publisher={IEEE}
}

@article{behnke2020boundary,
  title={Boundary conditions for traceability in food supply chains using blockchain technology},
  author={Behnke, Kay and Janssen, MFWHA},
  journal={International Journal of Information Management},
  volume={52},
  pages={101969},
  year={2020},
  publisher={Elsevier}
}

@article{behzadi2018agribusiness,
  title={Agribusiness supply chain risk management: A review of quantitative decision models},
  author={Behzadi, Golnar and O’Sullivan, Michael Justin and Olsen, Tava Lennon and Zhang, Abraham},
  journal={Omega},
  volume={79},
  pages={21--42},
  year={2018},
  publisher={Elsevier}
}

@article{barbosa2018opportunities,
  title={Opportunities and challenges in sustainable supply chain: An operations research perspective},
  author={Barbosa-P{\'o}voa, Ana Paula and da Silva, C{\'a}tia and Carvalho, Ana},
  journal={European journal of operational research},
  volume={268},
  number={2},
  pages={399--431},
  year={2018},
  publisher={Elsevier}
}

@article{treiblmaier2018impact,
  title={The impact of the blockchain on the supply chain: a theory-based research framework and a call for action},
  author={Treiblmaier, Horst},
  journal={Supply chain management: an international journal},
  volume={23},
  number={6},
  pages={545--559},
  year={2018},
  publisher={Emerald Publishing Limited}
}

@article{tonnissen2020analysing,
  title={Analysing the impact of blockchain-technology for operations and supply chain management: An explanatory model drawn from multiple case studies},
  author={T{\"o}nnissen, Stefan and Teuteberg, Frank},
  journal={International Journal of Information Management},
  volume={52},
  pages={101953},
  year={2020},
  publisher={Elsevier}
}

@article{casino2019systematic,
  title={A systematic literature review of blockchain-based applications: Current status, classification and open issues},
  author={Casino, Fran and Dasaklis, Thomas K and Patsakis, Constantinos},
  journal={Telematics and informatics},
  volume={36},
  pages={55--81},
  year={2019},
  publisher={Elsevier}
}

@article{feng2020applying,
  title={Applying blockchain technology to improve agri-food traceability: A review of development methods, benefits and challenges},
  author={Feng, Huanhuan and Wang, Xiang and Duan, Yanqing and Zhang, Jian and Zhang, Xiaoshuan},
  journal={Journal of cleaner production},
  volume={260},
  pages={121031},
  year={2020},
  publisher={Elsevier}
}

@article{al2019blockchain,
  title={Blockchain in industries: A survey},
  author={Al-Jaroodi, Jameela and Mohamed, Nader},
  journal={IEEE access},
  volume={7},
  pages={36500--36515},
  year={2019},
  publisher={IEEE}
}

@article{lim2021literature,
  title={A literature review of blockchain technology applications in supply chains: A comprehensive analysis of themes, methodologies and industries},
  author={Lim, Ming K and Li, Yan and Wang, Chao and Tseng, Ming-Lang},
  journal={Computers \& industrial engineering},
  volume={154},
  pages={107133},
  year={2021},
  publisher={Elsevier}
}

@article{azimifard2018selecting,
  title={Selecting sustainable supplier countries for Iran's steel industry at three levels by using AHP and TOPSIS methods},
  author={Azimifard, Arezoo and Moosavirad, Seyed Hamed and Ariafar, Shahram},
  journal={Resources Policy},
  volume={57},
  pages={30--44},
  year={2018},
  publisher={Elsevier}
}

@article{de2020blockchain,
  title={Blockchain and smart contracts in supply chain management: A game theoretic model},
  author={De Giovanni, Pietro},
  journal={International Journal of Production Economics},
  volume={228},
  pages={107855},
  year={2020},
  publisher={Elsevier}
}

@article{wang2020blockchain,
  title={Blockchain-enabled circular supply chain management: A system architecture for fast fashion},
  author={Wang, Bill and Luo, Wen and Zhang, Abraham and Tian, Zonggui and Li, Zhi},
  journal={Computers in Industry},
  volume={123},
  pages={103324},
  year={2020},
  publisher={Elsevier}
}

@article{singh2023revealing,
  title={Revealing the barriers of blockchain technology for supply chain transparency and sustainability in the construction industry: an application of pythagorean FAHP methods},
  author={Singh, Atul Kumar and Kumar, VR Prasath and Irfan, Muhammad and Mohandes, Saeed Reza and Awan, Usama},
  journal={Sustainability},
  volume={15},
  number={13},
  pages={10681},
  year={2023},
  publisher={MDPI}
}

@article{yang2022edge,
  title={Edge-cloud blockchain and IoE-enabled quality management platform for perishable supply chain logistics},
  author={Yang, Chen and Lan, Shulin and Zhao, Zhiheng and Zhang, Mengdi and Wu, Wei and Huang, George Q},
  journal={IEEE Internet of Things Journal},
  volume={10},
  number={4},
  pages={3264--3275},
  year={2022},
  publisher={IEEE}
}

@article{krinidis2010robust,
  title={A robust fuzzy local information C-means clustering algorithm},
  author={Krinidis, Stelios and Chatzis, Vassilios},
  journal={IEEE transactions on image processing},
  volume={19},
  number={5},
  pages={1328--1337},
  year={2010},
  publisher={IEEE}
}

@article{sremac2018anfis,
  title={ANFIS model for determining the economic order quantity},
  author={Sremac, Sini{\v{s}}a and Tanackov, Ilija and Kopi{\'c}, Milo{\v{s}} and Radovi{\'c}, Dunja},
  journal={Decision Making: Applications in Management and Engineering},
  volume={1},
  number={2},
  pages={81--92},
  year={2018}
}

@book{hyndman2018forecasting,
  title={Forecasting: principles and practice},
  author={Hyndman, RJ},
  year={2018},
  publisher={OTexts}
}

@article{mercier2018neural,
  title={Neural network models for predicting perishable food temperatures along the supply chain},
  author={Mercier, Samuel and Uysal, Ismail},
  journal={Biosystems engineering},
  volume={171},
  pages={91--100},
  year={2018},
  publisher={Elsevier}
}

@article{varriale2021sustainable,
  title={Sustainable supply chains with blockchain, IoT and RFID: A simulation on order management},
  author={Varriale, Vincenzo and Cammarano, Antonello and Michelino, Francesca and Caputo, Mauro},
  journal={Sustainability},
  volume={13},
  number={11},
  pages={6372},
  year={2021},
  publisher={MDPI}
}

@article{farhadi2023leveraging,
  title={Leveraging meta-learning to improve unsupervised domain adaptation},
  author={Farhadi, Amirfarhad and Sharifi, Arash},
  journal={The Computer Journal},
  pages={bxad104},
  year={2023},
  publisher={Oxford University Press}
}

@article{farhadi2024domain,
  title={Domain adaptation in reinforcement learning: a comprehensive and systematic study},
  author={Farhadi, A and Mirzarezaee, M and Sharifi, A and Teshnehlab, M},
  journal={Journal of Zhejiang University SCIENCE C},
  year={2024}
}

@article{farhadi2023unsupervised,
  title={Unsupervised Domain Adaptation for image classification based on Deep Neural Networks},
  author={Farhadi, Amirfarhad and Mirzarezaee, Mitra and Sharifi, Arash and Teshnehlab, Mohammad},
  journal={Intelligent Multimedia Processing and Communication Systems (IMPCS)},
  volume={4},
  number={1},
  pages={27--37},
  year={2023}
}

@inproceedings{mahalegi2024generative,
  title={Generative Adversarial Networks for Heterogeneous Unsupervised Domain Adaptation Detection},
  author={Mahalegi, Homayoun Safarpour Motealegh and Farhadi, Amirfarhad and Moln{\'a}r, Gy{\"o}rgy and Nagy, Enik{\H{o}}},
  booktitle={2024 IEEE 28th International Conference on Intelligent Engineering Systems (INES)},
  pages={000235--000244},
  year={2024},
  organization={IEEE}
}

@article{lataran2024developing,
  title={Developing a BERT-Enhanced Blockchain Model for HealthInsurance Fraud Detection},
  author={Lataran, Parham Najafi and Farhadi, Amirfarhad and Zamanifar, Azadeh and Taheri, Alireza},
  journal={Journal of Information Systems Research and Practice},
  volume={2},
  number={3},
  pages={56--61},
  year={2024}
}

@article{taheri2024enhancing,
  title={Enhancing aspect-based sentiment analysis using data augmentation based on back-translation},
  author={Taheri, Alireza and Zamanifar, Azadeh and Farhadi, Amirfarhad},
  journal={International Journal of Data Science and Analytics},
  pages={1--26},
  year={2024},
  publisher={Springer}
}
%
%
\end{document}